\documentclass[twocolumn,showpacs,preprintnumbers,amsmath,amssymb,prb]{revtex4}
\usepackage{graphicx}% Include figure files
\usepackage{dcolumn}% Align table columns on decimal point
\usepackage{bm}% bold math

\begin{document}

\preprint{APS/123-QED}

\title{Quadrupolar effect and rattling motion
\\in heavy fermion superconductor PrOs$_4$Sb$_{12}$}% Force line breaks with \\

\author{Terutaka Goto, Yuichi Nemoto, Kazuhiro Sakai, Takashi Yamaguchi,
\\Mitsuhiro Akatsu, Tatsuya Yanagisawa, Hirofumi Hazama}\author{Kei Onuki}%
\affiliation{%
Graduate School of Science and Technology, Niigata University,
Niigata 950-2181, Japan}%

\author{Hitoshi Sugawara, Hideyuki Sato}
\affiliation{
Department of Physics, Tokyo Metropolitan University, Minami-Ohsawa 1-1, Hachioji, Tokyo 192-0397, Japan% with \\
}%
\date{\today}% It is always \today, today,

\begin{abstract}
The elastic properties of a filled skutterudite PrOs$_4$Sb$_{12}$ with a heavy Fermion superconductivity
at $T_{\rm C}=1.85$ K have been investigated. The elastic softening of $(C_{11}-C_{12})/2$ and $C_{44}$ with
lowering temperature down to $T_{\rm C}$ indicates that the quadrupolar fluctuation due to the CEF state plays a role
for the Cooper paring in superconducting phase of PrOs$_4$Sb$_{12}$. A Debye-type dispersion in the elastic constants
around 30 K revealed a thermally activated $\Gamma_{23}$ rattling due to the off-center Pr-atom motion obeying
$\tau=\tau_0\exp(E/k_{\rm B}T)$ with an attempt time $\tau_0=8.8\times10^{-11}$ sec and an activation energy $E=168$ K.
It is remarkable that the charge fluctuation of the off-center motion with $\Gamma_{23}$ symmetry may mix with
the quadrupolar fluctuation and enhance the elastic softening of $(C_{11}-C_{12})/2$ just above $T_{\rm C}$.
\end{abstract}

\maketitle

The rare-earth cubic compounds based on Pr$^{3+}$ ions have received much attention because various unusual properties
are expected at low temperatures. The system with a non-Kramers $\Gamma_3$ doublet possessing two quadrupoles
$O_2^0=(2J_z^2-J_x^2-J_y^2)/\sqrt{3}$ and $O_2^2=J_x^2-J_y^2$ favors a quadrupole ordering. We refer the $\Gamma_3$
ground state systems as a metallic compound PrPb$_3$ showing the antiferro-quadrupole ordering at $T_Q= 0.4$ K \cite{1}
and a semiconductor PrPtBi undergoing ferro-quadrupole ordering at $T_Q=1.2$ K \cite{2}. PrSb is known as a singlet
ground state system \cite{3}. The elastic constant $(C_{11}-C_{12})/2$ is responsible for the quadrupolar susceptibility of
$O_2^0$ and $O_2^2$ with $\Gamma_3$ symmetry, while $C_{44}$ is for the susceptibility of
$O_{yz}=J_yJ_z+J_zJ_y$, $O_{zx}=J_zJ_x+J_xJ_z$, $O_{xy}=J_xJ_y+J_yJ_x$ with $\Gamma_5$ symmetry. The softening of
$(C_{11}-C_{12})/2$ and $C_{44}$ is a useful prove to clarify the quadrupolar effects of Pr-based compounds.

Recently, Bauer \textit{et al}. have found a new-type of the heavy Fermion superconductor in a filled skutterudite
PrOs$_4$Sb$_{12}$ with space group T$_{\rm h}^5$ (Im$\bar3$) \cite{4}. The heavy Fermion state with
a large specific heat coefficient $\gamma=750$ mJ/mol$\cdot{\rm K}^2$ of PrOs$_4$Sb$_{12}$ exhibits
the superconducting transition at $T_{\rm C}=1.85$ K associated with a large jump $\Delta{C/T}_{\rm C} \sim 500$
mJ/mol$\cdot{\rm K}^2$. A sign of the double transition in the specific heat has been found \cite{5}. The thermal
transport measurement in fields suggests the two distinct superconducting phases in PrOs$_4$Sb$_{12}$ \cite{6}.
The nuclear spin relaxation rate $1/T_1$ of Sb indicates unconventional superconductivity possessing neither
a coherence peak nor a $T^3$-power law \cite{7}. The muon spin relaxation in PrOs$_4$Sb$_{12}$ yields a penetration
depth indicating a new-type of energy gap \cite{8}. The odd-parity Cooper pairing mediated by the quadrupole fluctuation
is argued as unconventional heavy Fermion superconductivity in PrOs$_4$Sb$_{12}$ \cite{9}. Because the magnetic
susceptibility is rather silent to distinguish non-magnetic $\Gamma_{23}$ doublet from $\Gamma_1$ singlet,
it has not been settled whether the CEF ground state of PrOs$_4$Sb$_{12}$ is $\Gamma_{23}$ doublet or $\Gamma_1$
singlet \cite{4,10}. The measurement of $(C_{11}-C_{12})/2$ and $C_{44}$ responsible for the quadrpolar susceptibility in
PrOs$_4$Sb$_{12}$ is a central issue to clarify the CEF state and the interplay of the quadrupolar fluctuation to
the superconductivity in PrOs$_4$Sb$_{12}$.

The reduction of the thermal conductivity in filled skutterudites RM$_4$Sb$_{12}$ (R: La or Ce. M: Fe or Co) is caused by
a rattling motion due to a weakly bounded rare-earth ion in an oversized cage of Sb-icosahedron \cite{11}.
The filled skutterudites with the cage are favorable for the thermoelectric device possessing a high coefficient of
merit \cite{12}. The ultrasonic measurements are generally useful to observe the rattling motion or off-center tunneling
motion. We refer the rattling motion in clathrate materials Sr$_8$Ga$_{16}$Ge$_{30}$ \cite{13} and Ce$_3$Pd$_{20}$Ge$_6$
\cite{14}, and an off-center tunneling of OH-ion doped in NaCl \cite{15,16}. Recently, our experiment on $C_{44}$ in
La$_3$Pd$_{20}$Ge$_6$ revealed the ultrasonic dispersion around 20 K due to the rattling motion and elastic softening
below 3 K due to off-center tunneling motion in cage \cite{17}. It became to be sure that the rattling and tunneling are
common features of the clathrate compound with oversized cage. Quite recently a small off-center displacement of Pr-ion in
PrOs$_4$Sb$_{12}$ has been observed by X-ray absorption measurements \cite{18}. The rattling motion in the clathrate
compound PrOs$_4$Sb$_{12}$ with the heavy Fermion superconductivity has not been clarified yet.

The single crystal of PrOs$_4$Sb$_{12}$ with a length of 1.2 mm along the [110] direction for the present
ultrasonic measurements was grown by a flux method. The ultrasonic velocity {\it v} was detected by a phase
comparator based on a mixer technology. The piezoelectric LiNbO$_3$ transducers of {\it x}-cut and 36$^\circ${\it y}-cut
were used for the measurements of transverse and longitudinal ultrasonic waves, respectively. The elastic constant
$C=\rho{v^2}$ of PrOs$_4$Sb$_{12}$ with a lattice constant {\it a} = 0.930311 nm was estimated by the density
$\rho= 9.75$ g/cm$^3$. A $^3$He-evaporation fridge down to 500 mK was employed. 

Figure 1 shows temperature dependence of $(C_{11}-C_{12})/2$ of PrOs$_4$Sb$_{12}$, which was obtained by the transverse
wave of 17 MHz propagating along $\mbox{\boldmath $k$}=[110]$ with polarization $\mbox{\boldmath $u$}=[1\bar{1}0]$.
This $(C_{11}-C_{12})/2$ mode is associated with the elastic strain $\varepsilon_v=\varepsilon_{xx}-\varepsilon_{yy}$.
The increase of $(C_{11}-C_{12})/2$ around 30 K of Fig. 1 originates from the Debye-type dispersion, where the ultrasonic
wave frequency $\omega$ coincides with a relaxation time $\tau$ of the system as $\omega\tau=1$. A relatively large
lattice parameter {\it a}=0.930311 nm in PrOs$_4$Sb$_{12}$ may lead to the rattling motion of an off-center Pr-ion
in an oversized cage of Sb-icosahedron. The resonant scattering of the ultrasonic wave by the rattling motion of Pr-ion
over a potential hill brings about the Debye-type dispersion. The experimental determination of a relaxation time $\tau$
of the rattling is discussed latter.  

\begin{figure}
\includegraphics[width=0.9\linewidth]{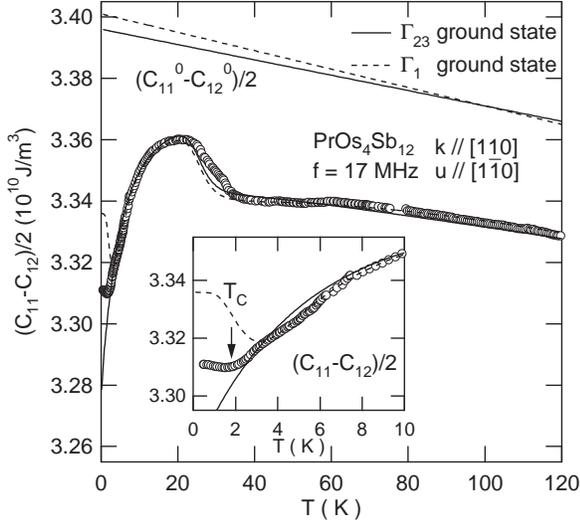}% Here is how to import EPS art
\caption{Temperature dependence of $(C_{11}-C_{12})/2$ in PrOs$_4$Sb$_{12}$ measured by ultrasonic wave of 17 MHz.
The anomaly around 30 K originates from the $\Gamma_{23}$ rattling due to the Pr-ion off-center motion. The softening of
$(C_{11}-C_{12})/2$ below 20 K down to superconducting point $T_{\rm C}=1.85$ K is due to the quadrupolar fluctuation of
the CEF states. The solid line and dashed line are fits by the quadrupolar susceptibility $\chi_Q$ for
$\Gamma_{23}$-$\Gamma_4^{(2)}$ and $\Gamma_1$-$\Gamma_4^{(2)}$ models, respectively.
Inset shows the detail around $T_{\rm C}$.}
\end{figure}

A remarkable softening of $(C_{11}-C_{12})/2$ below 20 K in Fig. 1 has been found with decreasing temperature.
As shown in inset of Fig. 1, the softening of $(C_{11}-C_{12})/2$ turns up around the superconducting transition
$T_{\rm C}=1.85$ K. The quadrupole-strain interaction,
$H_{\rm QS} = -\sum_{i}\sum_{\Gamma\gamma}g_{\Gamma}O_{\Gamma\gamma}(i)\varepsilon_{\Gamma\gamma}$,
and the inter-site quadrupole interaction,
$H_{\rm QQ} = -\sum_{i}g'_\Gamma \langle{O}_{\Gamma\gamma}\rangle{O}_{\Gamma\gamma}(i)$,
give rise to the elastic softening as
$C_\Gamma = C_\Gamma^0 - Ng_\Gamma^2\chi_{\rm Q}/(1-g'_\Gamma \chi_{Q})$ \cite{19}.
Here $\sum_i$ means a sum over rare earth sites in unit volume. $\chi_{\rm Q}$ is a quadrupolar susceptibility
consisting of the Curie-term for diagonal parts and the Van Vleck-term for off-diagonal ones. The coupling of
the quadrupole $O_2^2$ with $\Gamma_{23}$ symmetry to the elastic strain $\varepsilon_v$ is relevant for
the softening in $(C_{11}-C_{12})/2$ below 20 K of PrOs$_4$Sb$_{12}$.

The CEF Hamiltonian for the Pr$^{3+}$ ion with the site symmetry T$_{\rm h}$ is written as
$H_{CEF}=B_4O_4+B_6O_6+B_6^tO_6^t$, where $O_4=O_4^0+5O_4^4$, $O_6=O_6^0-21O_6^4$ and $O_6^t=O_6^2-O_6^6$ \cite{20}.
Two different types of the CEF models of $\Gamma_{23}$-$\Gamma_4^{(2)}$ and $\Gamma_1$-$\Gamma_4^{(2)}$ for
PrOs$_4$Sb$_{12}$ have been proposed so far \cite{4,21}. The solid line in Fig. 1 based on the doublet model of
$\Gamma_{23}$(0 K), $\Gamma_4^{(2)}$(8.2 K), $\Gamma_4^{(1)}$(133 K), and $\Gamma_1$(320 K) for
$B_4^0=6.75\times10^{-2}$ K, $B_6^0=-1.23\times10^{-3}$ K, and $B_6^t=-0.12\times10^{-2}$ K with
$|g_{\Gamma_{23}}|=97$ K and an inter-site coupling $g'_{\Gamma_{23}}=-0.27$ K reproduces the softening of
$(C_{11}-C_{12})/2$ mostly proportional to reciprocal temperature. The doublet model seems to be favorable for
the softening of $(C_{11}-C_{12})/2$. 

Recently, Kohgi et al. have proposed a singlet ground state CEF model of $\Gamma_1$(0 K), $\Gamma_4^{(2)}$(7.9 K),
$\Gamma_4^{(1)}$(135 K), and $\Gamma_{23}$(205 K) with $B_4^0=2.37\times10^{-2}$ K, $B_6^0=1.32\times10^{-3}$ K,
$B_6^t=1.08\times10^{-2}$ K \cite{21}. The dashed line for $|g_{\Gamma_{23}}|=79$ K and $g'_{\Gamma_{23}}=0.22$ K
based on the singlet model also reproduces the softening of $(C_{11}-C_{12})/2$ except for a small deviation below
a minimum around 3.5 K in the fitting. The off-center motion Pr ion with $\Gamma_{23}$-symmetry may enhance
the elastic softening of $(C_{11}-C_{12})/2$ just above $T_{\rm C}$, that will considerably renormalize the one-ion
susceptibility of the dashed line in Fig. 1.
Even in the case of the $\Gamma_{1}$-$\Gamma_{4}^{(2)}$ model, the charge fluctuation due to the off-center motion
may reproduce the softening of $(C_{11}-C_{12})/2$ just above $T_{\rm C}$ proportional to reciprocal temperature.
In order to settle the alternative CEF model of $\Gamma_{23}$-$\Gamma_{4}^{(2)}$ or $\Gamma_{1}$-$\Gamma_{4}^{(2)}$,
further experiments are necessary.
The Debye-type dispersion was employed to reproduce the anomaly around 20 K of solid and dashed lines in Fig. 1.
We discuss this point again in Fig. 3.

In Fig. 2, we show temperature dependence of $C_{44}$ obtained by the transverse wave propagating along [110] with
polarization along [001]. The softening of $C_{44}$ below 60 K is described in terms of the quadrupolar susceptibility
for the $\Gamma_4^{(2)}$-type quadrupole. The solid line in Fig. 2 is responsible for the $\Gamma_{23}$-$\Gamma_{4}^{(2)}$
model with parameters $|g_{\Gamma_4}|=34.4$ K and $g'_{\Gamma_4}=-0.002$ K. The dashed line is a fit for
the $\Gamma_{1}$-$\Gamma_4^{(2)}$ model with $|g_{\Gamma_4}|=70$ K and $g'_{\Gamma_4}=-0.07$ K. Because the quadrupolar
susceptibility of $C_{44}$ for both models is dominated by the Van Vleck term responsible for off-diagonal processes,
the determination of the CEF state by $C_{44}$ is rather indirect as similar as the magnetic susceptibility.
It should be noted that no sign of the ultrasonic dispersion has been found in $C_{44}$ around 30 K. 

\begin{figure}
\includegraphics[width=0.9\linewidth]{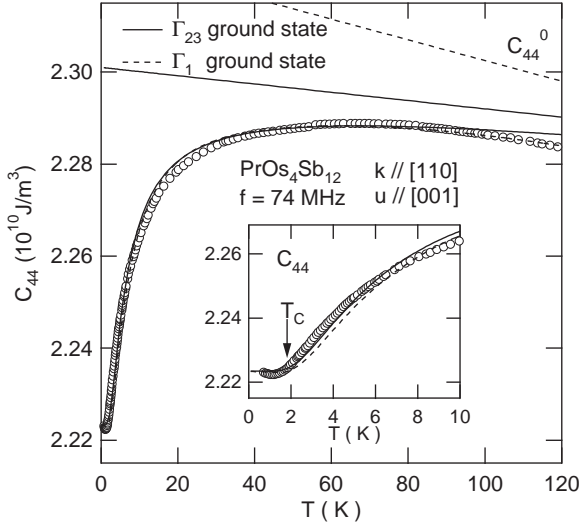}% Here is how to import EPS art
\caption{Temperature dependence of the elastic constant $C_{44}$ of PrOs$_4$Sb$_{12}$ measured by ultrasonic wave of 74 MHz.
The softening of below 60 K down to superconducting point $T_{\rm C}=1.85$ K is described in terms of the quadrupolar
susceptibility $\chi_Q$ of solid line for $\Gamma_{23}$-$\Gamma_4^{(2)}$ model and dashed line for
$\Gamma_1$-$\Gamma_4^{(2)}$ model. Inset shows the detail around $T_{\rm C}$.}
\end{figure}

Figure 3 represents $C_{\rm L}=(C_{11}+C_{12}+2C_{44})/2$ of PrOs$_4$Sb$_{12}$ obtained by the longitudinal wave
along the [110] direction. A remarkable frequency dependence of 17.8 MHz, 52.0 MHz and 87.7 MHz in Fig. 3 is described in
terms of the Debye-type dispersion as,
$C_{\rm L}(\omega)=C_{\rm L}(\infty)-\{C_{\rm L}(\infty)-C_{\rm L}(0)\}/(1+\omega^2\tau^2)$.
Here $\omega$ is the angular frequencies of the ultrasonic wave. Arrows in Fig. 3 indicate the temperature being
the resonant condition of $\omega\tau=1$. The anomaly of $(C_{11}-C_{12})/2$ around 30 K is also well described by
the Debye dispersion of the solid and dashed lines in Fig. 1.

In inset of Fig. 3 the temperature dependence of the relaxation time $\tau$ obtained by $C_{\rm L}$ is presented together
with the results of $(C_{11}-C_{12})/2$ of Fig. 1 and $C_{11}$, that is not presented here. The relaxation time
due to the rattling motion obeys the temperature dependence of $\tau=\tau_0\exp(E/k_{\rm B}T)$ with an attempt time
$\tau_0=8.8\times10^{-11}$ sec and an activation energy $E=168$ K. Utilizing a harmonic oscillation of
$\zeta(z)=(1/\pi{z}_0)^{1/2}\exp(-z^2/2z_0^2)$, we estimated a mean square displacement
$z_0=(1/ 2\pi)(h\tau_0/M)^{1/2}=0.079$ nm for Pr-ion in the present potential of the cage \cite{22}.
This result is comparable to $z_0=0.048$ nm of Ce$_3$Pd$_{20}$Ge$_6$ \cite{14} and $z_0=0.012$ nm
of La$_3$Pd$_{20}$Ge$_6$ \cite{17}.

The twelve Sb-atoms have a distance 0.3542 nm from the center of the cage. Because the Sb-atom is absent along
the [100] axis, the Pr-ion may favor an off-center motion over six minimum points of potential at
$r_1=(a,0,0)$, $r_2=(-a,0,0)$, $r_3=(0,a,0)$, $r_4=(0,-a,0)$, $r_5=(0,0,a)$, $r_6=(0,0,-a)$.
Here, the mean square displacement of Pr-atom extends over the potential minimums as $a\sim{z}_0/2=0.04$ nm.
On the other hand, the Os-atom locating at 0.4028 nm from the center of the cage along the three-fold [111] axis
prevents the off-center motion along the [111] axis. The Sb-atoms closely locating to the two-fold [110] axis
may hinder the off-center motion along the [110] axis.

\begin{figure}
\includegraphics[width=0.85\linewidth]{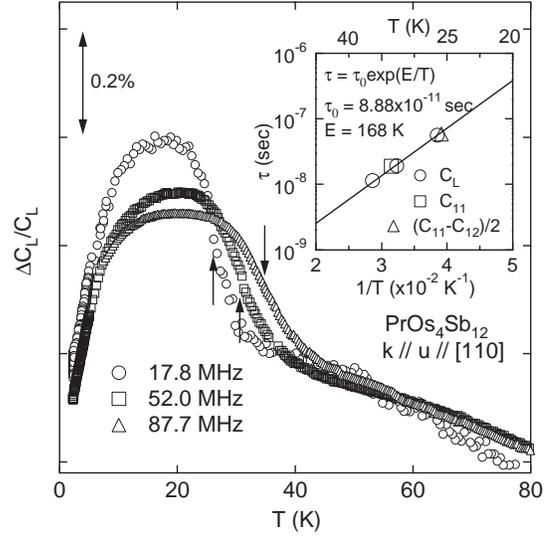}% Here is how to import EPS art
\caption{Temperature dependence of $C_{\rm L}=(C_{11}+C_{12}+2C_{44})/2$ in
PrOs$_4$Sb$_{12}$ measured by ultrasonic waves of 17.8, 52.0 and 87.7 MHz.
Arrows indicate the temperatures that the relaxation time $\tau$ of Pr-ion rattling coincides with the sound wave
frequencies $\omega$ as $\omega\tau=1$. Inset shows temperature dependence of relaxation time.}
\end{figure}

The ultrasonic dispersion has been observed in the transverse wave of $(C_{11}-C_{12})/2$ of the strain
$\varepsilon_v$ with $\Gamma_{23}$ symmetry. The longitudinal modes of $C_{11}$ and $C_{\rm L}$ consisting of
the strain $\varepsilon_u=(2\varepsilon_{zz}-\varepsilon_{xx}-\varepsilon_{yy})/\sqrt3$ with $\Gamma_{23}$ symmetry
in part also show ultrasonic dispersion. On the other hand, the $C_{44}$ mode responsible for the elastic strain with
$\Gamma_4^{(2)}$ symmetry does not show the dispersion effect. These results indicate that the thermally activated
rattling motion being coupled to the elastic strains $\varepsilon_u$ and $\varepsilon_v$ has the $\Gamma_{23}$ symmetry.
It is of particular importance to project out the off-center mode for the irreducible representations
at a center of the cage with the point group symmetry T$_{\rm h}$ \cite{23,24,25}.
Operating a symmetry element {\it R} of T$_{\rm h}$ on the atomic density $\rho_i=\rho(\rho_i)$ at the minimum point
$\rho_i(i = 1,2,\cdot\cdot\cdot,6)$, one obtains six dimensional representation matrices $D_{ij}(R)$.
The character $\chi(R)$ being a trace of the representation matrix reduces to direct sum
$\Gamma_1\oplus\Gamma_{23}\oplus\Gamma_4^{(2)}$.
The projection operator is used to pick up the $\Gamma_1$, $\Gamma_{23}$ and $\Gamma_4^{(2)}$
representations consisting of the fractional atomic density of Pr-ion over the six minimum points.
In the present case of PrOs$_4$Sb$_{12}$, the $\Gamma_{23}$ off-center mode of
$\rho_{\Gamma_{23u}}=2\rho_5+2\rho_6-\rho_1-\rho_2-\rho_3-\rho_4$ and
$\rho_{\Gamma_{23v}}=\rho_1+\rho_2-\rho_3-\rho_4$ with the fractional atomic distribution in Fig. 4
is the ground state of the system.
$\rho_{\Gamma_{23u}}$ means the distribution of fraction 1/2 at $\rho_5$ and $\rho_6$ sites and null
at $\rho_1,\rho_2,\rho_3,\rho_4$. And $\rho_{\Gamma_{23v}}$ is responsible for fraction 1/3 at
$\rho_1$ and $\rho_2$, 1/6 at $\rho_5$ and $\rho_6$, and null at $\rho_3$ and $\rho_4$. The total symmetric mode
$\rho_{\Gamma_1}=\rho_1+\rho_2+\rho_3+\rho_4+\rho_5+\rho_6$ and the polar mode
$\rho_{\Gamma_{4x}}=\rho_1-\rho_2$, $\rho_{\Gamma_{4y}}=\rho_3-\rho_4$, $\rho_{\Gamma_{4z}}=\rho_5-\rho_6$
may correspond to the excited states.
$\rho_{\Gamma_1}$ represents the mean fraction 1/6 over the six sites. $\rho_{\Gamma_{4x}}$,
for instance, has fraction 1/3 at $\rho_1$ and null at $\rho_2$, and fraction 1/6 at $\rho_3,\rho_4,\rho_5,\rho_6$. 

\begin{figure}
\includegraphics[width=0.9\linewidth]{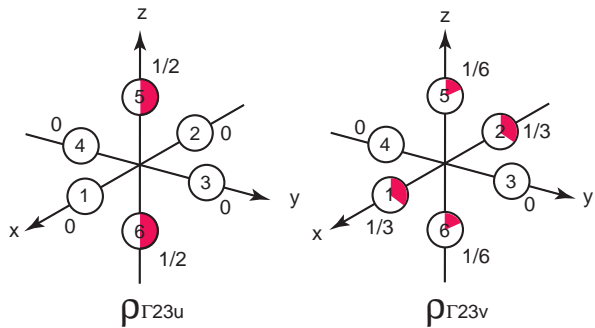}% Here is how to import EPS art
\caption{The $\Gamma_{23}$ rattling mode of 
$\rho_{\Gamma_{23u}}=2\rho_5+2\rho_6-\rho_1-\rho_2-\rho_3-\rho_4$
and $\rho_{\Gamma_{23v}}=\rho_1+\rho_2-\rho_3-\rho_4$ being responsible for
the ultrasonic dispersion in PrOs$_4$Sb$_{12}$.}
\end{figure}

Recent ultrasonic measurements on the clathrate compound La$_3$Pd$_{20}$Ge$_6$ by our group has successfully showed
the rattling and tunneling motions of an off-center La ion in cage \cite{17}. In the present clathrate compound of
PrOs$_4$Sb$_{12}$ with a cage of Sb-icosahedron, the thermally activated rattling motion over the potential hill
brings about the ultrasonic dispersion with the relaxation time $\tau$ in inset of Fig. 3. 
With lowering temperature, the thermally activated rattling dies out completely without showing the structural transition.
Consequently, the tunneling motion of Pr-ion through the hill in keeping the site symmetry to be T$_{\rm h}$
is relevant at low temperatures.
The tunneling motion being accompanied by charge fluctuation interacts with conduction electrons
through the channels of Pr-Sb bonding.
It is remarkable that the off-center tunneling motion with $\Gamma_{23}$ symmetry may bring
about the enhancement of the elastic softening in $(C_{11}-C_{12})/2$ just above $T_{\rm C}$.
The theoretical work for the interplay of the tunneling motion to the superconductivity
by Cox and Zawadowski \cite{22} may be relevant for the present PrOs$_4$Sb$_{12}$.

In conclusion we have successfully observed the elastic softening in $(C_{11}-C_{12})/2$ and $C_{44}$ above $T_{\rm C}$.
It is, however, still difficult to determine a CEF state in the alternative model of $\Gamma_{23}$-$\Gamma_4^{(2)}$ or
$\Gamma_1$-$\Gamma_4^{(2)}$. The field dependence of the elastic constant in particular is necessary to settle the CEF state
of the system. Nevertheless, it is worthwhile to emphasize the fact that the softening of $(C_{11}-C_{12})/2$ and $C_{44}$
indicate a crucial role of the quadrupolar fluctuation to the heavy Fermion superconductivity in PrOs$_4$Sb$_{12}$.
Furthermore, the ultrasonic dispersion due to the $\Gamma_{23}$ rattling motion with activation energy $E=168$ K has been
found. The $\Gamma_{23}$-type charge fluctuation associated with the off-center tunneling motion in particular may enhance
the elastic softening of $(C_{11}-C_{12})/2$ just above $T_{\rm C}$. The more accurate investigation is necessary to clarify
the CEF state in PrOs$_4$Sb$_{12}$ and the interplay of the quadrupole fluctuation and the off-center motion of Pr-ion to
the unconventional superconductivity.

The authors thank M.Kohgi, Y.Kuramoto, K.Miyake, and Y.$\bar{\rm O}$no for suggestive comments and discussions.
The present work was supported by a Grant-in-Aid for Scientific Research Priority Area "Skutterudite" (No.15072206)
of the Ministry of Education, Culture, Sports, Science and Technology, Japan.

\end{document}